\documentstyle[12pt]{article}
\def\fun#1#2{\lower3.6pt\vbox{\baselineskip0pt\lineskip.9pt
\ialign{$\mathsurround=0pt#1\hfil##\hfil$\crcr#2\crcr\sim\crcr}}}

\newcommand{\be}{\begin{eqnarray}}
\newcommand{\ee}{\end{eqnarray}}

\input{epsfig.sty}

\begin{document}

\Huge{\noindent{Istituto\\Nazionale\\Fisica\\Nucleare}}

\vspace{-3.9cm}

\Large{\rightline{Sezione SANIT\`{A}}}
\normalsize{}
\rightline{Istituto Superiore di Sanit\`{a}}
\rightline{Viale Regina Elena 299}
\rightline{I-00161 Roma, Italy}

\vspace{0.65cm}

\rightline{INFN-ISS 96/4}
\rightline{May 1996}

\vspace{2cm}

\begin{center}

\LARGE{Weak Decay Form Factors of Heavy Pseudoscalar Mesons
within a Light-Front Constituent Quark Model}\\

\vspace{1cm}

\large{I.L. Grach$^{(a)}$, I.M. Narodetskii$^{(a)}$, S. Simula$^{(b)}$}

\vspace{0.5cm}

\normalsize{$^{(a)}$Institute for Theoretical and Experimental Physics,
\\ 117259 Moscow, Russia \\ $^{(b)}$Istituto Nazionale di Fisica
Nucleare, Sezione Sanit\`{a}, \\ Viale Regina Elena 299, I-00161 Roma,
Italy}

\end{center}

\vspace{1cm}

\begin{abstract}

The transition form factors describing the semileptonic decays
of heavy pseudoscalar mesons are investigated within a relativistic
constituent quark model formulated on the light-front. For the first
time, the form factors are calculated in the whole accessible
kinematical region, adopting meson wave functions derived from various
effective $q \bar{q}$ interactions able to describe the meson mass
spectra. It is shown that the decay rates of the $B \to D \ell
\nu_{\ell}$, $D \to K \ell \nu_{\ell}$ and $D \to \pi \ell \nu_{\ell}$
weak decays are mainly governed by the effects of the confinement
scale, whereas the $B \to \pi \ell \nu_{\ell}$ decay rate is sensitive
to the high-momentum components generated in the meson wave functions
by the effective one-gluon-exchange interaction. Our results are
consistent with available experimental data and predictions of lattice
$QCD$ calculations and $QCD$ sum rules.  

\end{abstract}

\vspace{1cm}


\vspace{0.5cm}


\newpage

\pagestyle{plain}

\indent The investigation of semileptonic heavy-quark decays is an
important test of our understanding of weak and strong interactions.
As is well known, the semileptonic decay amplitude factorizes into the
product of the leptonic and hadronic $V-A$ currents. The matrix
elements of the latter contain relevant pieces of information about
the strong forces which bind quarks and gluons into hadrons, whereas
the leptonic part depends on the Cabibbo-Kobayashi-Maskawa ($CKM$)
mixing parameters, $V_{Q_1 Q_2}$, which are fundamental quantities of
the Standard Model. In order to calculate the decay rates and, hence,
to extract the $CKM$ parameters from the experiments, it is necessary
to know the hadron form factors in the whole region of accessible
values of the squared four-momentum transfer, $0 \leq q^2 \leq
q_{max}^2$. It should be pointed out that till now the predictions
obtained within various non-perturbative approaches, like, e.g., the
$QCD$ sum rule technique and the constituent quark ($CQ$) model, do
not cover the full range of $q^2$. In particular, as far as $CQ$
models are concerned, the weak hadron form factors are usually
calculated at a fixed value of $q^2$ (appropriate for the specific
$CQ$ model) and then extrapolated to the whole range of $q^2$. As a
matter of fact, in the original $WSB$ approach \cite{WSB} the form
factors are calculated at $q^2 = 0$ and then extrapolated to $q^2 > 0$
using a monopole ansatz $1 / (1 - q^2 / M_{pole}^2)$, with $M_{pole}$
being the mass of the lowest-lying vector meson in the given channel.
Within the $ISGW$ model \cite{ISGW} the form factors are calculated at
the point of zero recoil (i.e., $q^2 = q_{max}^2$) and then
extrapolated to $q^2 = 0$ using an exponential-like ansatz. The $CQ$
model of Ref. \cite{JAUS} uses the light-front ($LF$) formalism to
compute the form factors for space-like values of $q^2$; then, the
extrapolation to the time-like region is performed using a
two-parameter formula, which reproduces the values of the form factors
and their first two derivatives at $q^2 = 0$. Finally, in Ref.
\cite{MEL96} the form factors have been calculated in the space-like
region using a dispersion integral approach and, then, analytically
continuated in the time-like one. 

\indent In Ref. \cite{DGNS} the $q^2$ dependence of various
heavy-to-heavy and heavy-to-light form factors among pseudoscalar
mesons has been calculated in the whole accessible kinematical region.
The calculations have been based on a $LF$ $CQ$ model, adopting a
gaussian-like ansatz for the meson wave functions, and carried out in
a reference frame where the momentum transfer is purely longitudinal,
i.e. in a frame appropriate for time-like values of $q^2$. It has been
shown \cite{DGNS} that the time-like $LF$ result, obtained using the
matrix element of the "good" component of the weak vector current $J^+
= J^0 + J^3$, coincides with the contribution of the so-called
spectator pole of the Feynman triangle diagram. The remaining part of
the Feynman diagram, the so-called Z-graph, corresponds to the
contribution of quark-pair creation from the vacuum. The Z-graph
contribution is expected to be marginal for heavy-to-heavy decays (see
\cite{DGNS}) and to become more important near the zero-recoil point
for heavy-to-light decays (see \cite{IW90}). In this letter we use the
approach of Ref. \cite{DGNS}, but, as far as the wave functions of the
initial and final pseudoscalar mesons are concerned, we adopt the
eigenfunctions of $LF$ mass operators, constructed using various
effective $q \bar{q}$ interactions able to reproduce the meson mass
spectra. In this way a link between heavy-meson weak decay properties
and the "spectroscopic" constituent quark model is explicitly
constructed. It is shown that the decay rates for the $B \to D \ell
\nu_{\ell}$, $D \to K \ell \nu_{\ell}$ and $D \to \pi \ell \nu_{\ell}$
decays are mainly governed by the effects of the confinement scale,
whereas the $B \to \pi \ell \nu_{\ell}$ decay rate and, hence, the
extraction of $|V_{ub}|$ are sensitive to the high-momentum components
generated in the meson wave functions by the effective
one-gluon-exchange ($OGE$) interaction. Our results obtained both for
the form factors and the decay rates are compared with available
experimental data and predictions of various non-perturbative
approaches, like, e.g., lattice $QCD$ simulations and $QCD$ sum
rules.  

\indent Here and below we denote by $P_1$ ($P_2$) and $M_1$ ($M_2$) the
four-momenta and masses of the parent (daughter) mesons, respectively.
The momenta and masses of the corresponding active quarks will be
denoted as $p_1$ ($p_2$) and $m_1$ ($m_2$), whereas $k$ ($m$) is the
momentum (mass) of the light spectator (anti)quark. The four-momentum
transfer is $q = P_1 - P_2$ and the fraction $y$ is defined as $y =
P_2^+ / P_1^+$, where $P^+ \equiv P^0 + P^3$ with the 3-axis defining
the spin quantization axis. In a frame where the three-momentum
transfer is purely longitudinal (i.e, $\vec{q}_{\perp} = 0$), it can
be easily verified that: $q^2 = (1 - y) (M_1^2 - M_2^2 / y)$. Thus,
one gets: $y_{1,2} = \zeta (\eta \pm \sqrt{\eta^2 - 1})$, where $\zeta
= M_2 / M_1$ and $\eta = U_1 \cdot U_2$, with $U_1$ and $U_2$ being the
four-velocities of the parent and daughter mesons, respectively. The
two signs for $y_{1,2}$ correspond to whether the three-momentum
transfer is in the positive or negative direction of the 3-axis. At
the point of zero recoil, $q^2 = q^2_{max} = (M_1 - M_2)^2$, one has
$y_1 = y_2 = \zeta $, while $y_1 = 1$ and $y_2 = \zeta^2$ at $q^2 = 0$.

\indent The semileptonic decay of a pseudoscalar meson $Q_1 \bar{q}$
into another pseudoscalar meson $Q_2 \bar{q}$ is governed by the
weak vector current only. The transition form factors $h_{\pm}(q^2)$
and $f_{\pm}(q^2)$ are defined as
 \be
    <P_2| \bar{Q}_2 \gamma_{\mu} Q_1|P_1> & = & \sqrt{M_1 M_2} ~
    \left[ h_+(q^2) ~ (U_1 + U_2)_{\mu} + h_-(q^2) ~ (U_1 - U_2)_{\mu}
    \right] \nonumber \\
    & = & f_+(q^2) ~ (P_1 + P_2)_{\mu} + f_-(q^2) ~ (P_1 - P_2)_{\mu}
    \label{2}
 \ee
Neglecting lepton masses, the inclusive semileptonic decay rate
$\Gamma$ is given by
 \be
    \Gamma = |V_{Q_1Q_2}|^2 {G_F^2 M_1^5 \over 12 \pi^3 } \zeta^4
    \int_1^{\eta_{max}} d\eta ~ (\eta^2 - 1)^{3/2} ~ f_+^2(\eta)
    \label{4}
 \ee
where $\eta_{\max} = (\zeta + 1 / \zeta) / 2$ and $V_{Q_1Q_2}$ is the
relevant $CKM$ matrix element. The form factors $f_{\pm}(q^2)$ and
$h_{\pm}(q^2)$ can be evaluated using only the "good" component of the
current $J^+ = \bar{Q}_2 \gamma^+ Q_1$. The matrix elements of $J^+$
depend upon both $q^2$ and $y$, whereas the form factors $f_{\pm}$
($h_{\pm}$) are independent of $y$. In order to invert Eq. (\ref{2}),
two matrix elements $J^+(q^2, y) \equiv <P_2| \bar{Q}_2 \gamma^+
Q_1|P_1>$, corresponding to $y = y_1$ and $y = y_2$, should be
calculated, obtaining: $J^+(q^2, y_i) = 2P_1^+ ~ \left[ f_+(q^2) ~ (1
+ y_i) + f_-(q^2) ~ (1 - y_i) \right]$ with $i = 1, 2$. Putting
$H(q^2, y_i) \equiv J^+(q^2, y_i) / 2P^+_1$ one has
 \be
    f_{\pm}(q^2) = { (y_2 \mp 1) H(q^2, y_1) - (y_1 \mp 1) H(q^2, y_2)
    \over y_2 - y_1}
    \label{3}
 \ee
Following Ref. \cite{DGNS}, the $LF$ ground state of the parent meson
$Q_1 \bar{q}$ can be written as
 \be
    |P_1> = \sum_{\lambda \bar{\lambda}} \int {d\vec{k}_{\perp} \over
    \sqrt{2 (2 \pi)^3}} {dx \over \sqrt{x(1 - x)}} ~ \sqrt{A_1(x,
    k_{\perp}^2)} ~ R^{(1)}_{00}(x, \vec{k}_{\perp}; \lambda
    \bar{\lambda}) ~ {w_1(k^2) \over \sqrt{4 \pi}} ~ Q_1^{\dagger}
    \bar{q}^{\dagger} |0>
    \label{5}
 \ee
where $Q_1^{\dagger} \equiv Q_1^{\dagger}(\vec{p}_1, \lambda) $ and
$\bar{q}^{\dagger} \equiv \bar{q}^{\dagger}(\vec{k}, \bar{\lambda})$
are the creation operators of the heavy active quark and the light
spectator antiquark, with helicities $\lambda$ and $\bar{\lambda}$,
respectively. The $LF$ momenta are taken as: $\vec{k} = (k^+,
\vec{k}_{\perp})$ and $\vec{p}_1 \equiv \vec{P}_1 - \vec{k} = (P_1^+ -
k^+, -\vec{k}_{\perp})$, where $\vec{k}_{\perp}$ is the relative
transverse momentum. The spectator and the active quarks carry the
fractions $x$ and $1 - x$ of the plus component of the momentum of the
meson, respectively. In Eq. (\ref{5}) $R^{(1)}_{00}(x,
\vec{k}_{\perp}; \lambda \bar{\lambda})$ is the momentum-dependent
spin part of the meson wave function, arising from the Melosh
rotations of the $CQ$ spins (see \cite{DGNS}), and $A_1(x,
k_{\perp}^2) \equiv M_{10} [1 - (m_1^2 - m^2)^2 / M_{10}^4] ~ / ~ 4x(1
- x)$, with $M_{10}^2 = (m_1^2 + k_{\perp}^2) / (1 - x) + (m^2 +
k_{\perp}^2) / x$ being the squared free mass. Finally, $k^2 \equiv
k_{\perp}^2 + k_3^2$, where $k_3 \equiv (x - 1/2) M_{10} + (m_1^2 -
m^2) / 2M_{10}$. The state vector of the daughter meson has a form
analogous to (\ref{5}) with the obvious replacement $1 \leftrightarrow
2$. The states $|P_i>$ ($i = 1, 2$) are normalized as: $<P'_i|P_i> =
(2 \pi)^3 ~ 2 P_i^+ ~ \delta(P_i^+ - {P'}_i^+)$, so that the
normalization condition for $w_i(k^2)$ is $\int_0^{\infty} dk ~ k^2
w_i^2(k^2) = 1$. Putting $x' \equiv x / y$ for the $LF$ momentum
fraction of the spectator quark in the final meson, the matrix elements
$H(q^2, y)$ can be cast in the following form \cite{DGNS}
 \be
    H(q^2, y) & = & \int_0^y dx \int d\vec{k}_{\perp} ~ \sqrt{A_1(x,
    k_{\perp}^2) ~ A_2(x', k_{\perp}^2)} ~ {w_1(k^2) w_2(k'^2) \over 4
    \pi} \nonumber \\
    & & { [m (1 - x) + x m_1] ~ [m (1 - x') + x' m_2] + k_{\perp}^2
    \over \sqrt{[m (1 - x) + x m_1]^2 + k_{\perp}^2} ~ \sqrt{[m (1 -
    x') + x' m_2]^2 + k_{\perp}^2}}
    \label{6}
 \ee

\indent The radial wave function $w_i(k^2)$ can be chosen to be the
eigenfunction of an equal-time effective $Q \bar{q}$ Hamiltonian (see
for more details \cite{CAR}). For the latter we will consider two
choices: the first one corresponds to the relativized Godfrey-Isgur
($GI$) Hamiltonian \cite{GI85} and the second one is the
non-relativistic ($NR$) model of Ref. \cite{NCS92}. We want to point
out that both potential models nicely reproduce the mass spectra of
light as well as heavy mesons. In particular, the interaction terms
contain an effective $OGE$ interaction, composed by a spin-dependent
part, responsible for the hyperfine mass splitting in light mesons,
and a spin-independent term, responsible for the hydrogen-like pattern
of heavy-meson (bottonium) states. For comparison, we consider also a
Gaussian-like ansatz, with the harmonic oscillator ($HO$) length taken
from the updated version of the $ISGW$ model \cite{ISGW2}. In what
follows, the three choices will be referred to as the $GI$, $NR$ and
$HO$ cases, respectively. The $CQ$ momentum distribution corresponding
to the wave functions $w^{(GI)}$, $w^{(NR)}$ and $w^{(HO)}$, obtained
for the $D$-meson, is shown in Fig. 1. It can be seen that: i)
$w^{(HO)}(k^2)$ is quite similar to the wave function obtained by
retaining only the confining part of the $GI$ interaction, i.e. it
takes into account only the effects of the confinement scale; ii) both
$w^{(GI)}$ and $w^{(NR)}$ exhibit high-momentum components generated
by the effective $OGE$ term; iii) the high-momentum tail is higher for
the $GI$ interaction. Similar results (cf. \cite{CAR}) hold as well in
case of the other mesons ($\pi$, $K$ and $B$) considered in this
letter.

\indent The results obtained for the form factor $f_+(q^2)$ (Eqs.
(\ref{3}) and (\ref{6})) and the decay rate $\Gamma$ (Eq. (\ref{4})) of
the $B \to D \ell \nu_{\ell}$, $B \to \pi \ell \nu_{\ell}$, $D \to K
\ell \nu_{\ell}$ and $D \to \pi \ell \nu_{\ell}$ decays, are reported
in Tables 1-3 and Figs. 2-3, and compared with results from lattice
$QCD$ calculations, $QCD$ sum rules and quark models. In our
calculations the $PDG$ \cite{PDG} values of the meson masses have been
adopted for all the three models of the meson wave function.

\indent {\bf Decay  $B \to D \ell \nu_{\ell}$}. From Table 1 it can
be seen that our results for $f_+^{B \to D}(q^2 = 0)$, $f_+^{B \to
D}(q_{max}^2)$ and the decay rate $\Gamma(B \to D \ell \nu_{\ell})$
depend only slightly upon the choice of the meson wave function, i.e.
these quantities are mainly governed by the effects of the confinement
scale. Our calculated rates are in agreement with existing $CQ$ model
predictions, which typically range from $\sim 7$ to $\sim 13 ~
ps^{-1}$. By combining the branching ratio $Br(B^0 \to D^- \ell^+
\nu_{\ell})  = (1.9 \pm 0.5) \%$ \cite{PDG} with the world average of
the $B^0$ lifetime, $\tau_{B^0} = 1.50 \pm 0.11 ~ ps$ \cite{PDG}, the
measured rate is $\Gamma (B^0 \to D^- \ell^+ \nu_{\ell}) = (1.27 \pm
0.35) \cdot 10^{10} ~ s^{-1} $. From our predicted rates (see Table
1), one gets: $|V_{bc}| = 0.034 \pm 0.005$ ($GI$) and $|V_{bc}| =
0.036 \pm 0.005$ ($NR$ and $HO$). These predictions, which are
expected to be modified only slightly by radiative corrections (see
\cite{NEU94}), are in agreement with the updated "experimental"
determinations of $|V_{bc}|$ obtained from exclusive and inclusive
semileptonic decays of the B-meson, viz. $|V_{bc}|_{excl} = 0.0373 \pm
0.0045_{exp} \pm 0.0065_{th}$ and $|V_{bc}|_{incl} = 0.0398 \pm
0.0008_{exp} \pm 0.0040_{th}$ \cite{SKW95}. At the point of zero
recoil the form factor $h_+(\eta = 1)$ results to be: $0.960$ ($GI$),
$0.971$ ($NR$), $0.965$ ($HO$), in nice agreement with the updated
$QCD$ sum rule estimate, $1 + \delta_{1/m^2} = 0.945 \pm 0.025$
\cite{NEU94}. Moreover, the values of the slope, $\hat{\rho}^2$, and
the curvature, $\hat{c}$, of the form factor $h_+(\eta)$ at the point
of zero recoil, i.e. $h_+(\eta) \approx h_+(1) \{1 - \hat{\rho}^2
(\eta - 1) + \hat{c} (\eta - 1)^2) + O[(\eta -1)^3]\}\}$, are:
$\hat{\rho}^2 = 0.96 ~ (GI), 1.22 ~ (NR), 1.30 ~ (HO)$ and $\hat{c} =
0.71 ~ (GI), 0.96 ~ (NR), 1.20 ~ (HO)$. The results corresponding to
the $GI$ wave function compare very favourably with the recent
model-independent improved bounds obtained in \cite{CN95}. The full
$q^2$ dependence of the form factor $f^{B \to D}_+(q^2)$ is shown in
Fig. 2(a). It can be seen that: i) in the whole range of $q^2$ the
form factor $f_+^{B \to D}(q^2)$ is affected only slightly by the
high-momentum tail of the $B$- and $D$-meson wave functions; ii) the
monopole approximation, with $M_{pole} = 6.25 ~ GeV$ \cite{WSB},
yields a slightly flatter $q^2$ dependence.

\indent {\bf Decay $B \to \pi \ell \nu_{\ell}$}. The $q^2$ dependence
of the form factor $f_+^{B \to \pi}(q^2)$ is shown in Fig. 2(b). Near
the zero-recoil point our result does not exhibit any pole dominance
and deviates from recent results of lattice $QCD$ simulations
\cite{APE95}. Such a discrepancy can be ascribed to the effects
related to the admixture of the $|B^* \pi>$ component in the $B$-meson
wave function \cite{IW90}, which, in the $LF$ language, corresponds to
the contribution of the Z-graph (quark-pair creation diagram). We want
to point out that the decay rate $\Gamma(B \to \pi \ell \nu_{\ell})$
is not expected to be sharply affected by the Z-graph, because in Eq.
(\ref{4}) the contribution arising from the region near the
zero-recoil point is kinematically suppressed. From Table 1 it can
clearly be seen that the form factor $f^{B \to \pi}_+(q^2 = 0)$ and,
hence, the decay rate $\Gamma(B \to \pi \ell \nu_{\ell})$ are sensitive
to the high-momentum components generated in the meson wave functions
by the effective $OGE$ interaction, particularly in case of the $GI$
model. This feature is related to the huge value of the maximum recoil
achieved in the $B \to \pi \ell \nu_{\ell}$ transition. Moreover, our
$LF$ results are in overall agreement with the ones obtained from
recent lattice $QCD$ simulations \cite{APE95,UKQCD95}, and only
slightly lower than the results of the analysis performed in Ref.
\cite{KR95} using the $LF$ $QCD$ sum rule (see Tables 2-3). Recently,
the $CLEO$ collaboration \cite{CLEO_1} has reported the first signal
for exclusive semileptonic decays of the $B$ meson into charmless final
states, in particular for the decay channel $B \to \pi \ell
\nu_{\ell}$.  Adopting the models of Refs. \cite{WSB} and \cite{ISGW}
to reconstruct the efficencies, the "model-dependent" branching ratios
are: $Br(B \to \pi \ell \nu_{\ell}) = (1.63 \pm 0.57) \cdot 10^{-4}$
and $(1.34 \pm 0.45) \cdot 10^{-4}$, respectively. Combining the
average of these results with the world average value of the $B^0$
lifetime, one gets: $\Gamma (B^0 \to \pi^- \ell^+ \nu_{\ell}) = (0.99
\pm 0.25) \cdot 10^{-4} ~ ps^{-1}$. From our predicted rates (see
Table 1) one has: $|V_{bu}| = 0.0025 \pm 0.0003$ ($GI$), $0.0030 \pm
0.0004$ ($NR$) and $0.0032 \pm 0.0004$ ($HO$). This means that the
extraction of $|V_{ub}|$ is sensitive to the effects due to the
high-momentum tail of the meson wave functions. We stress that this
conclusion is not expected to be sharply modified by the effects
related to the quark-pair creation from the vacuum. Using the results
previously obtained for $|V_{bc}|$, one has: $|V_{bu} / V_{cb}| =
0.074 \pm 0.014$ ($GI$), $0.083 \pm 0.016$ ($NR$) and $0.088 \pm
0.016$ ($HO$). These results are in accord with the value of $|V_{bu} /
V_{cb}|$ derived from measurements of the end-point region of the
lepton spectrum in inclusive semileptonic decays \cite{ARGUS,CLEO_2},
viz. $|V_{bu} / V_{cb}|_{incl} = 0.08 \pm 0.01_{exp} \pm 0.02_{th}$.

\indent {\bf Decay $D \to K \ell \nu_{\ell}$}. The $q^2$ behaviour of
the form factor $f^{D \to K}_+(q^2)$ is shown in Fig. 3(a). It can be
seen that our form factor is mainly governed by the effects of the
confinement scale and it is consistent both with the assumption of pole
dominance and with the results of the lattice $QCD$ calculations of
Refs. \cite{APE95,BER91}. Moreover, our results for $f^{D \to K}_+(0)$
and $f^{D \to K}_+(q^2_{max})$ (see Table 1) compare favourably with
the recent experimental results \cite{FAB95} $f^{D \to K}_+(0) = 0.77
\pm 0.01 \pm 0.04$ and $f^{D \to K}_+(q^2_{max}) = 1.42 \pm 0.25$,
obtained from the measured lepton spectrum assuming pole dominance. By
combining the experimental data on the branching ratio, $Br(D^0 \to
K^- e^+ \nu_e) = 3.68 \pm 0.21 \%$ \cite{PDG}, with the accurate value
of the $D^0$ lifetime, $\tau(D^0) = 0.415 \pm 0.004 ~ ps$ \cite{PDG},
one has: $\Gamma (D^0 \to K^- e^+ \nu_e) = 0.089 \pm 0.005 ~ ps^{-1}$.
From our predicted rates (see Table 1) one obtains: $|V_{cs}| = 0.88
\pm 0.03$ ($GI$), $0.90 \pm 0.03$ ($NR$) and $0.89 \pm 0.03$ ($HO$).
The constraint of unitarity of the $CKM$ matrix with three generations
of leptons yields a ($\sim 10 \%$) higher value, viz. $|V_{cs}| =
0.974$ \cite{PDG}.

\indent {\bf Decay $D \to \pi \ell \nu_{\ell}$}. This is the only
heavy-to-light decay where an extensive comparison with experiment is
possible. Our results for the form factor $f_+^{D \to \pi}(q^2)$ are
shown in Fig. 3(b). It can be seen that its behaviour at low $q^2$
agrees with the prediction  of the $LF$ $QCD$ sum rule of Ref.
\cite{KR95}, whereas near $q^2 = q_{max}^2$ it deviates from the pole
approximation. The same discussion, already done about the relevance
of the Z-graph in the $B \to \pi$ decay, applies as well to the $D \to
\pi$ transition. From Tables 2-3 it can be seen that our values for
$f_+^{D \to \pi}(q^2 = 0)$ and $\Gamma (D \to \pi \ell \nu_{\ell})$ are
in nice agreement both with the experimental data \cite{PDG} and the
results obtained within various non-perturbative approaches. Assuming
$|V_{cd}| = 0.221 \pm 0.003$ \cite{PDG} from the unitarity of the
$CKM$ matrix, our predictions for the semileptonic decay rate are:
$\Gamma (D^0 \to \pi^- e^+ \nu_e) = 8.2 \pm 0.2$ ($GI$), $7.6 \pm 0.2$
($NR$) and $7.8 \pm 0.2$ ($HO$), in units $10^{-3} ~ ps^{-1}$. These
results compares favourably with the experimental value
$(9.4_{-2.9}^{+5.5}) \cdot 10^{-3} ~ ps^{-1}$ \cite{PDG} and the
recent $LF$ $QCD$ sum rule prediction $(7.6 \pm 0.2) \cdot 10^{-3} ~
ps^{-1} $ \cite{KR95}. The ratio of the branching fraction of the
Cabibbo suppressed decay $D \to \pi \ell \nu_{\ell}$ to that of the
Cabibbo favoured decay $D \to K \ell \nu_{\ell}$ has been recently
determined by the CLEO collaboration by measuring both charged
\cite{CLEO_3} and neutral \cite{CLEO_4} $D$-meson decays. Assuming the
pole dominance for the form factors, the following values have been
obtained: $|f^{D \to \pi}_+(0) / f^{D \to K}_+(0)| = 1.29 \pm 0.21 \pm
0.11$ \cite{CLEO_3} and $1.01 \pm 0.20 \pm 0.07$ \cite{CLEO_4}. With
respect to the average of these values we predict a slightly lower
ratio, viz. $|f^{D \to \pi}_+(0) / f^{D \to K}_+(0)| = 0.91$ ($GI$),
$0.88$ ($NR$) and $0.87$ ($HO$). The value obtained in Ref.
\cite{ISGW2} is $0.71$ and other theoretical predictions typically
range from $0.7$ to $1.4$.

\indent In conclusion, adopting a light-front constituent quark model
we have investigated the transition form factors, which govern
the heavy-to-heavy and heavy-to-light semileptonic weak decays between
pseudoscalar mesons. For the first time, the form factors have been
calculated in the whole kinematical region accessible in semileptonic
decays, adopting meson wave functions derived from various effective
$q \bar{q}$ interactions able to describe the meson mass spectra. It
has been shown that the decay rates for the $B \to D \ell \nu_{\ell}$,
$D \to K \ell \nu_{\ell}$ and $D \to \pi \ell \nu_{\ell}$ weak decays
are mainly governed by the effects of the confinement scale, whereas
the $B \to \pi \ell \nu_{\ell}$ decay rate and, hence, the extraction
of $|V_{ub}|$ are sensitive to the high-momentum components generated
in the meson wave functions by the effective one-gluon-exchange
interaction. Our results both for the form factor and the decay rates
are consistent with available experimental data and predictions of
lattice $QCD$ calculations and $QCD$ sum rules. We want to stress that
an estimate of the contribution of quark-pair creation from the vacuum
is mandatory, particularly in case of the $B \to \pi \ell \nu_{\ell}$
and $D \to \pi \ell \nu_{\ell}$ transitions.

\vspace{0.5cm}

We are very grateful to K.A. Ter-Martirosyan for valuable discussions.
Two of us (I.L.G. and I.M.N.) acknowledge the financial support of the
INTAS grant No 93-0079. This work was done in part under the RFFR
grant, Ref. No. 95-02-04808a.

\newpage

\section*{Table Captions}

\begin{description}

\item[Table 1.] The form factor $f_+(q^2)$ (Eqs. (\ref{3}) and
(\ref{6}))  for various semileptonic decays, evaluated at $q^2 = 0$ and
$q^2 = q^2_{\max}$ using the $GI$, $NR$ and $HO$ wave functions (see
text), and the corresponding decay rate $\Gamma$ (Eq. (\ref{4})),
calculated in units $ps^{-1}$ assuming $|V_{Q_1 Q_2}| = 1$.

\item[Table 2.] The form factor $f_+(q^2)$ of the $B \to \pi \ell
\nu_{\ell}$ and $D \to \pi \ell \nu_{\ell}$ transitions evaluated at
$q^2 = 0$ within different approaches. The label $Exp$ means the
experimental result (assuming pole dominance), while the labels $SR$
and $LAT$ correspond to $QCD$ sum rule and lattice $QCD$ calculations,
respectively.

\item[Table 3.] The same as in Table 2, but for the decay rate $\Gamma$
of the $B \to \pi \ell \nu_{\ell}$ and $D \to \pi \ell \nu_{\ell}$
transitions, calculated in units $10^{13}s^{-1}$ and $10^{11}s^{-1}$,
respectively, assuming $|V_{Q_1 Q_2}| = 1$. The experimental result
quoted for the $D \to \pi \ell \nu_{\ell}$ transition has been taken
from \cite{PDG} assuming $|V_{cd}| = 0.221$. The results from the
lattice $QCD$ simulations of Refs. \cite{APE95,UKQCD95} are obtained
assuming pole dominance.

\end{description}

\vspace{1cm}

\section*{Figure Captions}

\begin{description}

\item[Fig. 1.] The $CQ$ momentum distribution $|w(k^2)|^2$ versus the
internal momentum $k$ in the $D$-meson. The solid, dashed and dotted
lines correspond to the $GI$, $NR$ and $HO$ cases,  respectively (see
text). The dot-dashed line is the result obtained using the
eigenfunction of the Hamiltonian of Ref. \cite{GI85} in which only the
confining part of the interaction term is retained.

\item[Fig. 2.] The form factor $f_+(q^2)$ of the $B \to D \ell
\nu_{\ell}$ (a) and $B \to \pi \ell \nu_{\ell}$ (b) transitions. The
solid, dashed and dotted lines correspond to the results of our $LF$
calculations (Eq. (\ref{6})), obtained using the $GI$, $NR$ and $HO$
wave functions, respectively. The dot-dashed lines are the monopole
approximation $f_+(q^2) = f_+(0) / (1 - q^2 / M_{pole}^2)$ with
$M_{pole} = 6.25 ~ GeV$ (a) and $5.33 ~ GeV$ (b). In (b) the open dots
and squares are the $QCD$ sum rule predictions of Refs. \cite{KR95} and
\cite{BAL91}, respectively, whereas the full dots are the lattice
$QCD$ results from Ref. \cite{APE95}.

\item[Fig. 3.] The same as in Fig. 2, but for the $D \to K \ell
\nu_{\ell}$ (a) and $D \to \pi \ell \nu_{\ell}$ (b) transitions. The
dot-dashed lines are the monopole  approximation with $M_{pole} = 2.11
~ GeV$ (a) and $2.01 ~ GeV$ (b). The full triangles are the predictions
of the lattice $QCD$ simulations of Ref. \cite{BER91}. 

\end{description}

\newpage

\centerline{\bf Table 1}

\vspace{0.5cm}

\begin{center}

\begin{tabular}{||c|c|c|c|c||} \hline
weak       & wave     & $f_+(0)$ & $f_+(q^2_{\max})$ & $\Gamma$ \\
transition & function &          &                   &          \\
\hline

$B \to D$   & GI & 0.765 & 1.294 & 11.13  \\
            & NR & 0.692 & 1.322 & 9.835  \\
            & HO & 0.684 & 1.365 & 9.780  \\ \hline

$B \to \pi$ & GI & 0.464 & 1.803 & 15.21  \\
            & NR & 0.361 & 1.524 & 11.12  \\
            & HO & 0.293 & 1.658 &  9.624 \\ \hline

$D \to K$   & GI & 0.835 & 1.379 & 0.114  \\
            & NR & 0.787 & 1.598 & 0.111  \\
            & HO & 0.780 & 1.560 & 0.112  \\ \hline

$D \to \pi$ & GI & 0.762 & 1.281 & 0.167  \\
            & NR & 0.694 & 1.216 & 0.156  \\
            & HO & 0.681 & 1.289 & 0.160  \\ \hline
\end{tabular}

\end{center}

\vspace{3cm}

\centerline{\bf Table 2}

\vspace{0.5cm}

\begin{center}
\begin{tabular}{||c|c|c||} \hline
& $f_+^{B \to \pi}(0)$ & $f^{D \to \pi}_+(0)$ \\ \hline

GI & 0.464 & 0.762 \\

NR & 0.361 & 0.694 \\

HO & 0.293 & 0.684 \\ \hline

SR ~ \cite{KR95}     & $0.29 \pm 0.01$ & $0.66 \pm 0.03$ \\

SR ~ \cite{BAL91}    & $0.26 \pm 0.02$ & $0.5  \pm 0.1 $ \\

SR ~ \cite{DP88}     & $0.4 \pm 0.1  $ & $0.75 \pm 0.05$ \\

WSB ~ \cite{WSB}     & $0.33         $ & $0.69$ \\

ISGW ~ \cite{ISGW}   & $0.09         $ & $0.51$ \\

LAT ~ \cite{APE95}   & $0.35 \pm 0.08$ & $0.80 \pm 0.08$ \\

LAT ~ \cite{UKQCD95} & $0.43 \pm 0.02$ & \\
Exp ~ \cite{PDG}     &                 & $0.75_{-0.15}^{+0.23} \pm
0.06$ \\ \hline \end{tabular}

\end{center}

\newpage

\centerline{\bf Table 3}

\vspace{0.5cm}

\begin{center}
\begin{tabular}{||c|c|c||}\hline
& $\Gamma(\bar{B}^0 \to \pi^+ e^- \bar{\nu})$ & $\Gamma(D^0 \to \pi^-
e^+ \nu)$ \\ \hline

GI & 1.521 & 1.67 \\
NR & 1.112 & 1.56 \\
HO & 0.962 & 1.60 \\ \hline

SR ~ \cite{KR95}     & $0.81$          & $1.56$ \\

SR ~ \cite{BAL91}    & $0.51 \pm 0.11$ & $0.80 \pm 0.17$ \\

SR ~ \cite{DP88}     & $1.45 \pm 0.59$ & $1.66_{-0.21}^{+0.23}$ \\

WSB ~ \cite{WSB}     & $0.74$          & $1.41$ \\

ISGW ~ \cite{ISGW}   & $0.21$          & $0.77$ \\

LAT ~ \cite{APE95}   & $0.8  \pm 0.4 $ & $1.8_{-0.4}^{+0.6}$ \\

LAT ~ \cite{UKQCD95} & $1.02 \pm 0.36$ & \\
Exp ~ \cite{PDG}     &                 & $1.9_{-0.6}^{+1.1}$ \\ \hline
\end{tabular}

\end{center}

\newpage

\begin{figure}

\epsfig{file=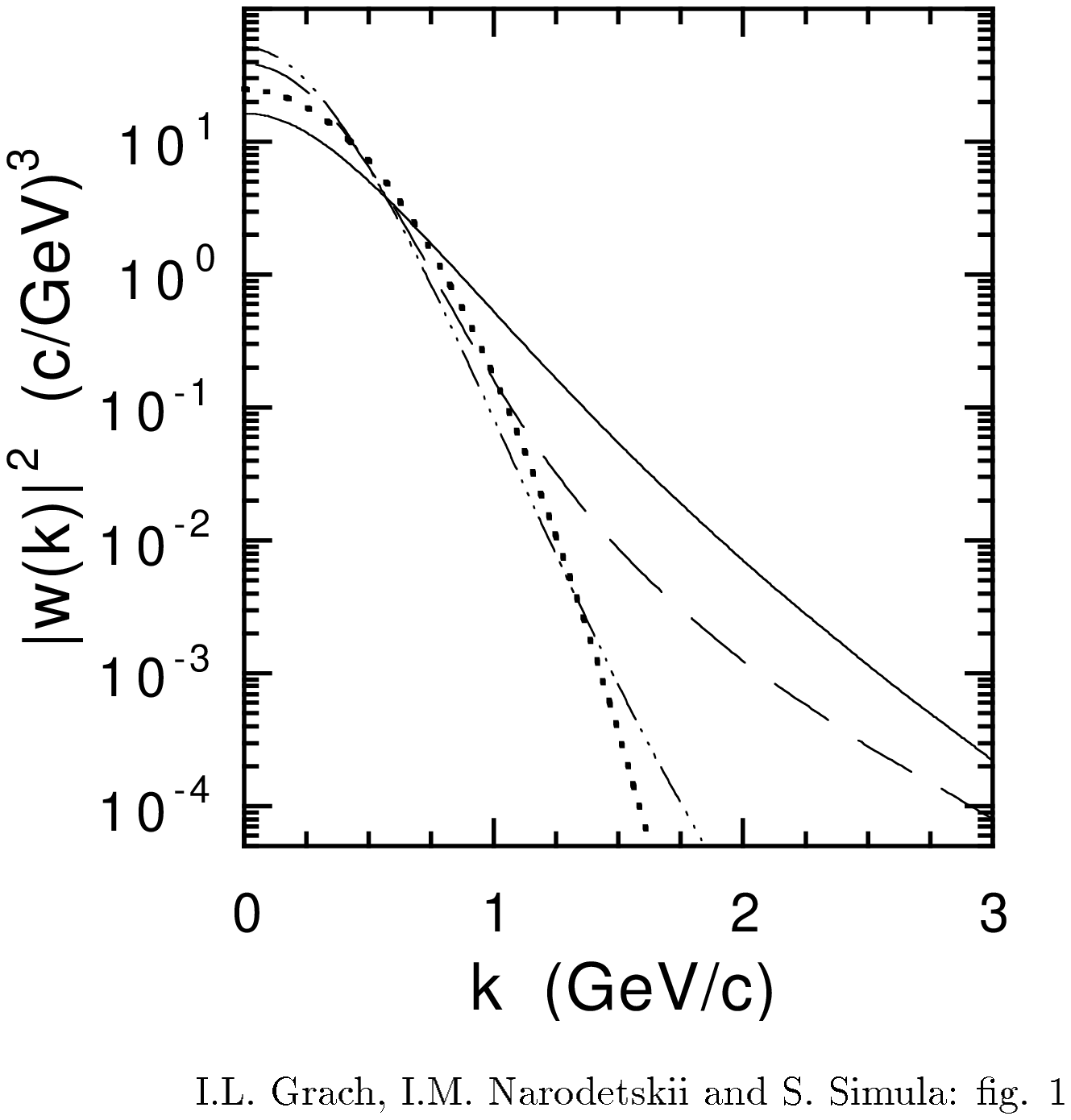}

\end{figure}

\newpage

\begin{figure}

\epsfig{file=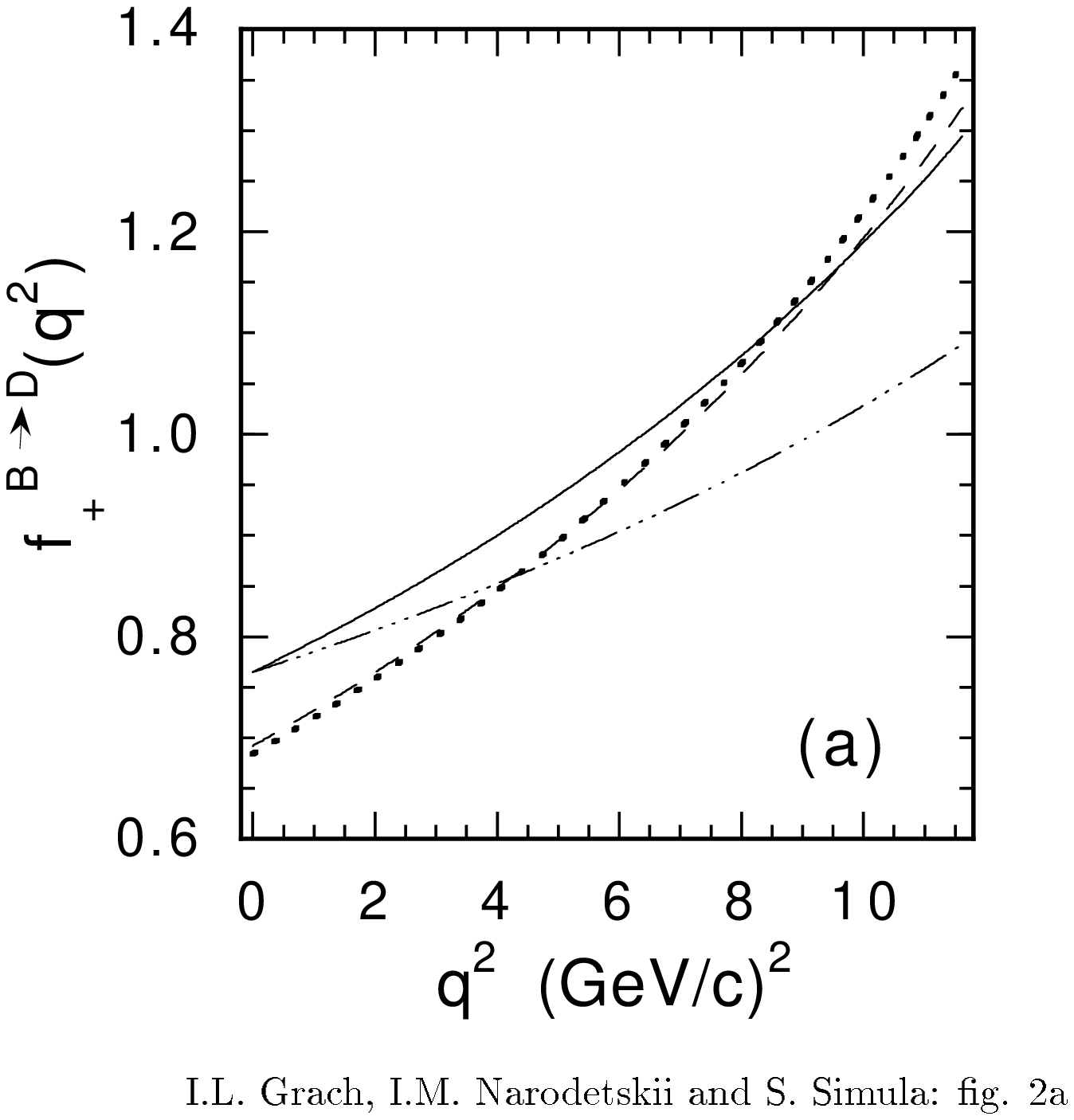}

\end{figure}

\newpage

\begin{figure}

\epsfig{file=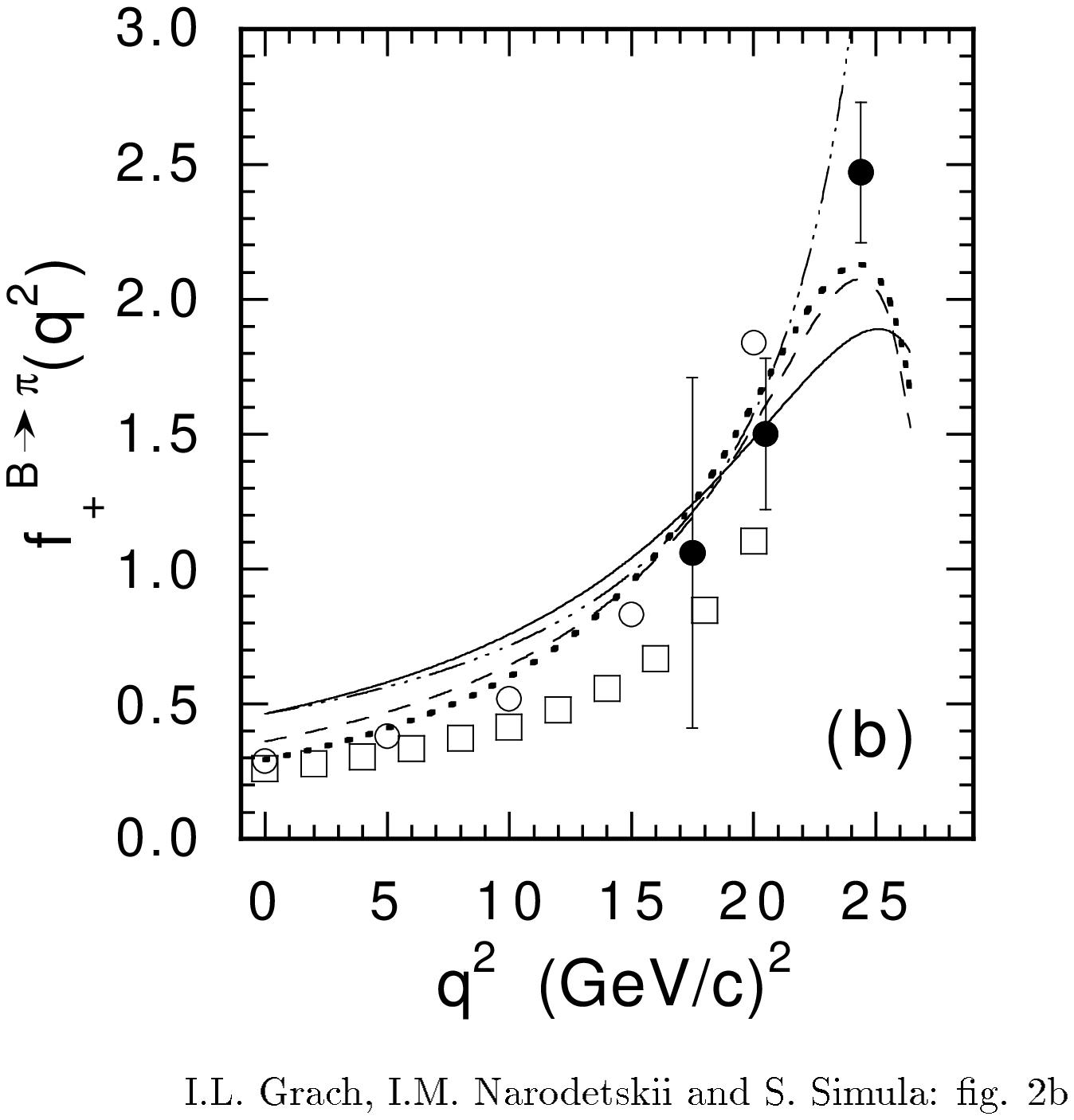}

\end{figure}

\newpage

\begin{figure}

\epsfig{file=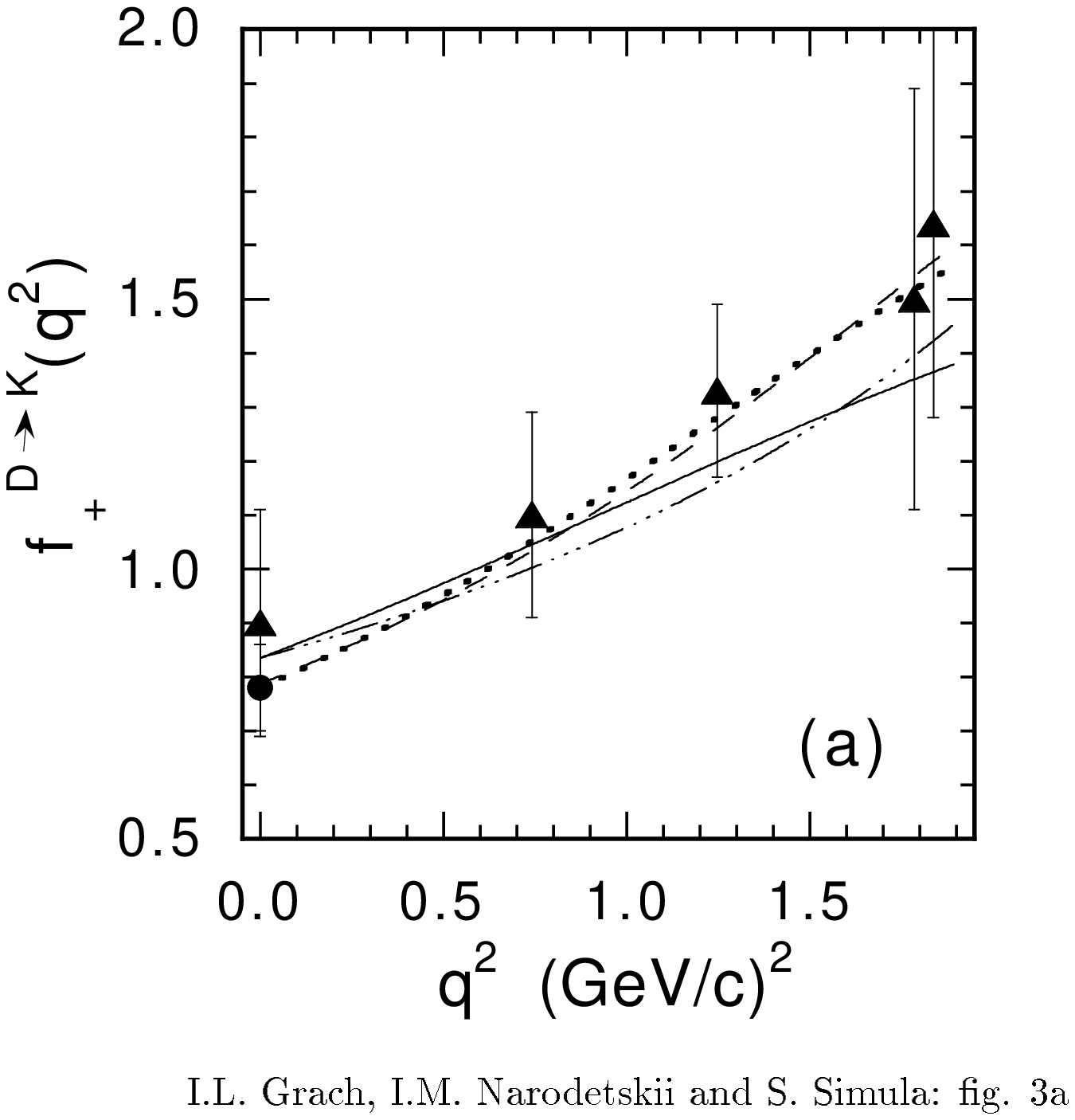}

\end{figure}

\newpage

\begin{figure}

\epsfig{file=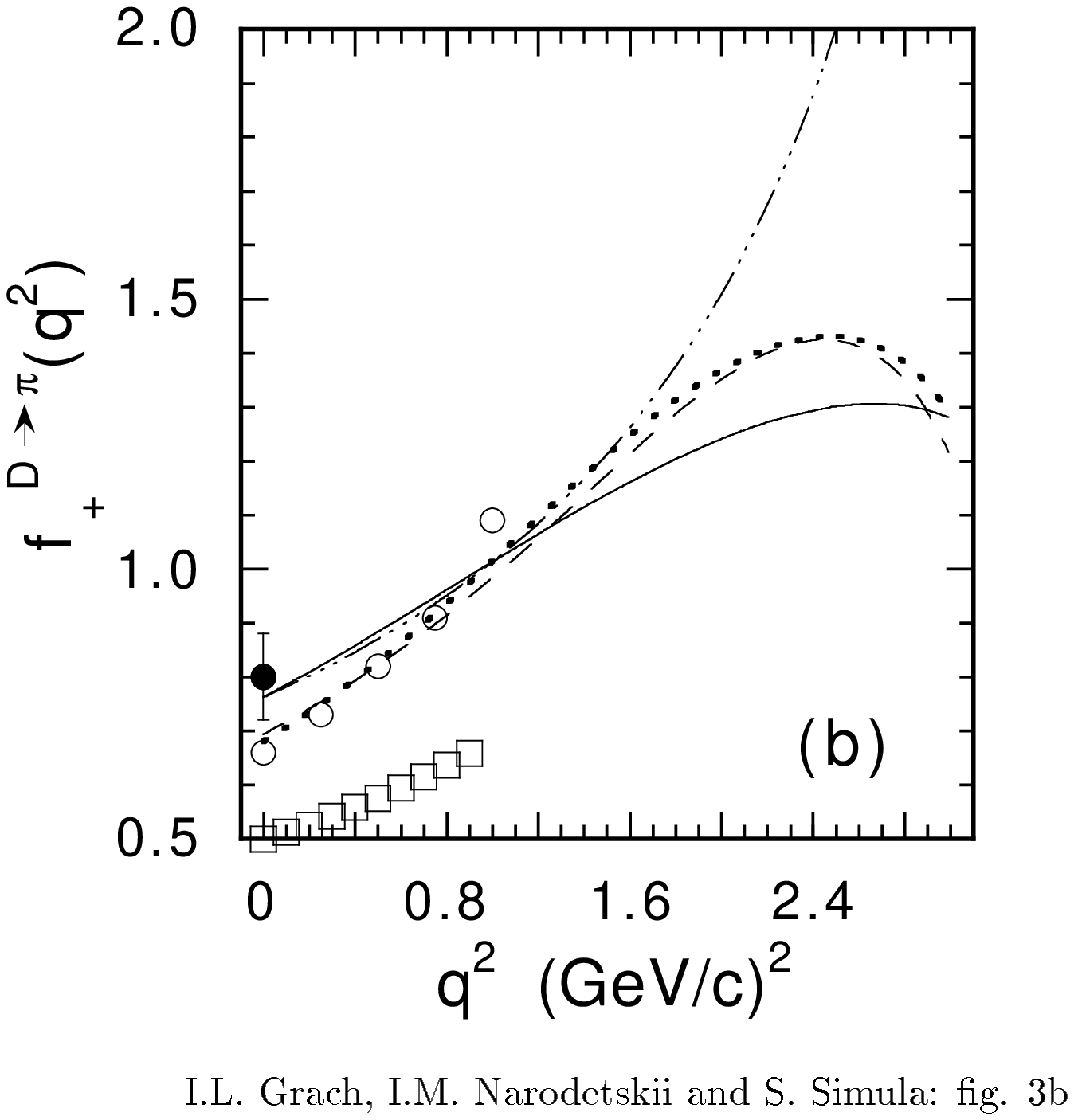}

\end{figure}

\end{document}